# Vulnerability Detection Approaches on Application Behaviors in Mobile Environment

Abdellah OUAGUID
RITM Laboratory, ESTC, Hassan II University, Casablanca, Morocco
ouaguid@gmail.com

Mohamed OUZZIF
RITM Laboratory, ESTC, Hassan II University, Casablanca, Morocco
ouzzif@gmail.com

Noreddine ABGHOUR
LIMSAD, FSAC, Hassan II University, Casablanca, Morocco
nabghour@gmail.com

## Abstract

Several solutions ensuring the dynamic detection of malicious activities on Android ecosystem have been proposed. These are represented by generic rules and models that identify any purported malicious behavior. However, the approaches adopted are far from being effective in detecting malware (listed or not) and whose form and behavior are likely to be different depending on the execution environment or the design of the malware itself (polymorphic for example). An additional difficulty is added when these approaches are unable to capture, analyze, and classify all the execution paths incorporated in the analyzed application earlier. This suggests that the functionality of the analyzed application can constitute a potential risk but never explored or revealed.
We have studied some malware detection techniques based on behavioral analysis of applications. The description, characteristics, and results obtained from each technique are presented in this article wherein we have also highlighted some open problems, challenges as well as the different possible future directions of research concerning behavioral analysis of malware.

## Introduction

The year 2019 has witnesses yet again Android dominance by 72.2% (StatCounter Global Stats 2020) of smart-phones worldwide market share. The large adoption of Android sets it as the preferred target of mobile cybercrime (Figure 1).

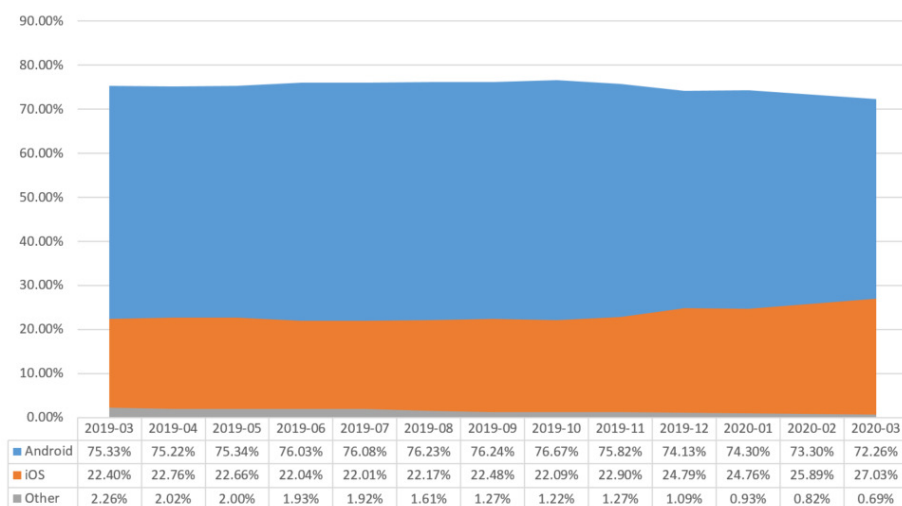

**Fig. 1: Mobile Operating System Market Share Worldwide (2020)**

Victim of its popularity, Android is an opportunity for hackers to get their hands on a large amount of victim data (SMS, GPS location, etc.) with less effort by taking advantage of free access to the source code of Android (known under the name AOSP). This makes installing and spreading malware easier compared to a closed operating system (iOS, BlackBerry, etc.).
The Android ecosystem also plays a role in the emergence of new vulnerabilities and the spread of hacker attack vectors. Indeed, several mobile manufacturers (Smartphone or Tablet) manage to design and market several products adapted to the needs and financial capacity of the consumer. These products do not host Android as published by Google but adopt a modified version including new proprietary components essential for its proper functioning (device driver for example) or presentation layers to stand out from the competition.

Specific and personalized developments made by manufacturers, and which are subsequently integrated into Android OS, may also contain vulnerabilities whose impact is of serious comparable to those found in applications installed by the user (Wu et al. 2013) or even in AOSP. The detected vulnerabilities affect the preinstalled components of specific manufacturers or specific ranges for a given manufacturer (Farhang, Kirdan, et al. 2020) (Elsabagh et al. 2020). This makes the Android ecosystem and its security segmented more and more (within the same manufacturers), and it only increases the risks associated with any equipment running Android such as Farhang, Laszka, and Grossklags (2018) shown in their study. With the evolution of mobile attacks, which have become increasingly stealthy (McAfee 2020), any flaws introduced by the manufacturer or developers of third-party applications can constitute a gateway for hackers to gain access to sensitive resources and data of hardware and the user by acquiring system privileges and permissions in an illegal manner, and without the consent of the end-user.

To address the various threats, in particular those caused by malware or those intentionally introduced into applications installed by the manufacturers or by the user, several analysis methods and approaches have been proposed by Kouliaridis et al. (2020) and others. In this work, we will focus on existing detection approaches based on malware behavioral analysis and the potential they have for identifying and detecting malicious behaviors that malicious applications may carry once installed in an adequate environment to start their malicious mission.

The rest of this article is organized as follows: Part 2 presents the different techniques for detecting vulnerabilities. Part 3 describes in detail the approach to behavioral malware analysis and the various solutions proposed. Part 4 presents possible future research prospects regarding the behavioral analysis of malware. Finally, Part 5 concludes the article.

## Vulnerability Detection Techniques

Several detection approaches and techniques have been proposed to ensure efficiency in the various phases of the analysis and the detection process to decide whether the application analyzed is malicious or not. To identify the functionality and how the malware activates, executes, and manipulates its data or stolen data, two main analysis techniques are used: static and dynamic analysis.

### *Static Code Analysis*

Static analysis of an application consists of reconstructing, scanning, and analyzing the code of the application without executing it. Its advantage is that it tries to cover and explore all possible paths of control and execution and examine the functionalities programmed without invoking them in order to extract, identify and analyze any malicious or legitimate action but which could be used maliciously by malware (system and API call (Aafer, Du, and Yin 2013), permission (Saracino et al. 2016), etc.).

All these processing can consume a lot of time and require resources in terms of calculation and storage without forgetting that the executable of the application must first undergo a treatment of "Reverse Engineering (RE)" (Eilam 2011), the latter is essential to have exploitable source code. However, RE is becoming increasingly unreliable (easily bypassed) and expensive because of the obfuscation techniques (Dalla Preda and Maggi 2017) (Suarez-Tangil et al. 2017) that the creator of malware adopts when compiling its source code. In several studies (those based on Machine Learning, such as Xi et al. (2019)), the static characteristics extracted from the analyzed applications which will be normalized, processed are used as input data to machine learning algorithms to decide whether the application is malicious or benign.

### *Dynamic Behavior Analysis*

Unlike static analysis, dynamic analysis of an application consists of observing and studying the behavior and actions taken by this application during its execution. It consists of listing and examining all activities related to the file system as Cabau, Buhu, and Oprisa (2016) proposed, to the network (Nari and Ghorbani 2013), to the sequences of API and system calls (functions, parameters, instructions, associated data flows) (Martinelli, Marulli, and Mercaldo 2017) (Egele et al. 2008), etc. Dynamic analysis is generally carried out in a controlled and virtual test environment where several monitoring software components are installed. This customization of the test environment can influence the normal and natural execution of malicious applications because it can generate a completely different behavior compared to a similar execution in a real environment. This will weaken the effectiveness of dynamic analysis in reducing the probability of the malware detection since the verification of rules and models already classified will be based on system calls and data generated by artificial behaviors that do not reflect the reality. Dynamic analysis requires continuous monitoring, adaptation, and updating for effective automation in the analysis or classification of malicious behaviors to follow the continuous evolution of the techniques used in malware. The complexity and the transformation capacity of a certain type of these malware (Polymorphic and Metamorphic malware) allow it to escape from the various traditional detection methods, thanks to the self-adaptation of malware functionalities according to the environment in which they are executed.

There are also hybrid detection techniques that try to merge the strengths of static and dynamic analysis (Anderson, Storlie, and Lane 2012) to improve the assessment and classification of the applications analyzed based on the collection and analysis of the functionalities detected via machine learning. Other approaches monitor the application during its execution to report any malicious behavior that may occur in the Smartphone (Faiella et al. 2017).

**Analysis of Behavior Detection Techniques**

Because of the integration of advanced techniques in the design and development of malwares (Wazid, Zeadally, and Das 2019), the detection of the latter is now more and more difficult despite the various works done in this direction. Its algorithms can be effective in specific cases for a context Well-defined. However, analyzing the behavior of malware in a real execution environment can be an important factor in identifying and detecting the impacts associated with the malicious behavior of malware and its future targets.

For an effective analysis of the integrated vulnerabilities in the applications or in AOSP itself, implicitly or explicitly (security flaws in the code), the behavioral analysis of the launched processes must be implemented and remain in active listening in order to control the various flows and interactions between its processes and the resources (of which they have access or not) from Smartphones. The data provided by these can contain information on behaviors that can be considered inappropriate or malicious, even if the resources used can already be made available to the applications, the source of their behaviors, thanks to the permissions granted upon installation or they are, in one way or another, put within its reach because of a security breach (following an elevation of privileges for example).

*Techniques for Detecting and Analyzing Malware*

To minimize human intervention in understanding and analyzing new malware, it is essential to automate its tasks in order to "robotize" their detections so that the potential risks associated with their malicious behavior can easily be identified. Despite the increase, diversification, and complication of the technics of malicious applications, the analysis, extraction, and sample classification processes of these applications must be capable of extracting models or images representing the identity of its malware. Its representations can then be used for the classification of any new application whose behavior is likely to harm the security and normal functioning of the end user's devices.

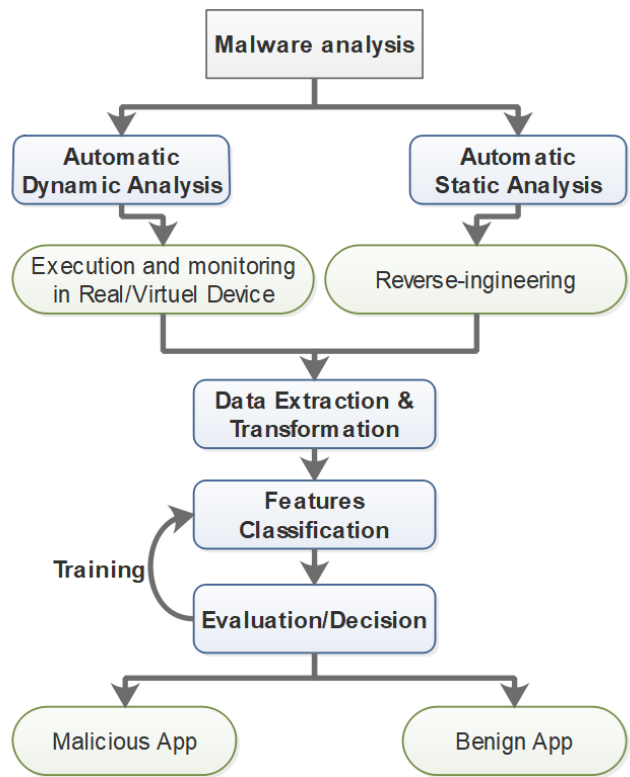

**Fig. 2: Malware detection schema**

The detection of a malicious application is mainly based on four steps (Figure 2):

- Analysis of the application: after receiving the executable of the application to be analyzed, the system begins the analysis step, which can be dynamic, static, or both:
  - Dynamic analysis: consists of the execution and monitoring of the application in a real or virtual machine. The observation of the activities of the application can relate to the launched processes, the network requests, the modifications made on the system files, the registry, etc.
  - Static analysis: consists of applying "Reverse-engineering" to the application to be able to decompile the executable and reconstruct its source code so that it can be used in the following steps.
- Data Extraction and Transformation: data mining techniques (Graph model, n-tuple, n-gram, etc.) can be used in this step to extract data that may contain characteristics, specifications, and functionalities (unitary or combined). This is geared towards determining the benign behaviors (representing the normal functioning of applications) or malicious behavior which could constitute a potential risk for the security of system resources and the integrity of sensitive user data. The extracted data can undergo transformations in order to eliminate unnecessary data (noise-tolerant) and then generate signatures, behavioral models, graphs representing the image of the extracted functionalities.
- Feature Classification: uses the Machine Learning to classify the characteristics of the analyzed application by referring to the generic rules and models previously built based on the characteristics already extracted from the application samples (benign and malicious) during the training phase of the system.
- Evaluation/Decision: in this step, the characteristics extracted from the analyzed application are compared with the signatures or behavioral model previously-stored the system database, and following this comparison, the system can deduce whether the analyzed application is malicious or not.

*Malware Behavior Detection Approach*

Several studies have been elaborated to describe the different malware detection approaches based on their behavioral imprint extracted during their interaction with the environment on which they are hosted. The nature of the latter, which can be a real environment, Sandbox, or a virtual machine, and its architecture play a decisive role in identifying the different application execution paths (malicious or not).

*1) MADAM:*
Saracino et al. (2016) proposed Madam, which is an Android malware detection system based on behavior monitoring at the level of the user (their activity/inactivity), the application (critical APIs), the package (permissions requested), and the kernel (system calls, etc.). After having classified most of the behavioral characteristics of real and known malware in misbehavior classes, the system ensures the extraction, analysis, and correlation of the functionalities of installed applications with behavioral models, previously generated, in order to be able to effectively detect and block any malicious activity. This system has shown its effectiveness in detecting more than 96% of malicious applications with a low rate of false alarms.

*2) M0droid:*
This technique about detecting Android malware had been proposed by Damshenas et al. (2015). It consists of two elements: a light client agent and an analyzer server. The client agent is installed in mobile devices that communicate with an analysis server, the latter analyzes the behavior of Android applications (invoked system calls) and creates signatures representing the malicious activities that malware can carry out; its signatures are used by client agents to identify and detect any vulnerability threatening users. The client-server architecture of this framework allows the end-user to minimize the consumption of resources and phone energy. The latter runs the client agent in the background and transmits to the analyzer server the various data collected regarding the installed applications. The analyzer server takes care of the analysis of the applications in an emulator by injecting it with random events both to invoke the system calls corresponding to the processing, and to cover the maximum of the possible paths of execution of the analyzed application. The behavior of the latter is captured and then used to create a standardized signature that will then be classified and compared with the blacklist of signatures that are previously constructed by this framework. Outsourcing the detection functionality to the cloud overcomes the limitations related to the performance of the end user's mobile device. 60.16% is the malware detection rate with 0.4% false negatives and 39.43% as a false positive rate.

*3) DroidDetector:*
The functionalities of dynamic and static analysis of Android applications have been combined in Droiddetector (Yuan, Lu, and Xue 2016), this hybrid solution is based on a Deep Learning method based on the Deep Belief Network (DBN) to extract, investigate and classify static characteristics (calls system, permissions, ...) and dynamics (behavior categories) of benign and malicious applications analyzed. The result of these treatments is used to feed the DBN network so that it can be used afterward by the DroidDetector mobile application previously installed on the end user's device. With a sample of 1,760 malicious applications, 192 static and dynamic characteristics were extracted, which allowed the level of detection accuracy to reach 96.76%.

*4) Andromaly:*
Andromaly (Shabtai et al. 2012) is a Host-based malware detection system that continuously monitors and classifies the various events dynamically generated by the mobile device while applications are running. Other features (related to network traffic, battery level, number of running processes, etc.) are also monitored by this system. Using Machine Learning algorithms, the behavioral characteristics collected are classified according to their nature (malicious or not), and then used to assess the level of infection of the mobile device. The alert history is also used and combined with the results obtained to reduce the rate of the false positives that are obtained. Even if the experiments are not done on samples of real malicious applications, the obtained results confirm that their solution is effective on the applications analyzed.

*5) ServiceMonitor:*

With the system proposed by Salehi, Amini, and Crispo (2019), malicious behaviors are identified using Machine Learning algorithms. Indeed, ServiceMonitor consists of generating a precise representation of the behavior of applications based on its different interactions with the service and system resources. The use of these resources is classified and represented by a statistical model (in the form of Markov chains) making it possible to monitor their exploitation by the various running applications. With a sample of 18,058 applications processed (10,024 benign and 8,034 malicious), ServiceMonitor was efficient and precise in detecting malicious behavior with an accuracy rate of 96.7%.

*6) **HIDROID**:*

Ribeiro et al. (2020) developed a novel Host-based IDPS for Android (HIDROID), which is a standalone Intrusion Detection and Prevention Systems that runs entirely on a mobile device without the need for a remote server. The behavior of applications with the device's resources (battery, CPU, memory, etc.) is monitored using a detection engine, which collects data in real-time and builds up a model representing benign behaviors by using a combination of statistics and Machine Learning algorithms that will help detect vulnerabilities (known or not). Any suspicious behavior is communicated to the user in the form of an alert, and countermeasures are taken to reduce the risk of an attack or intrusion. HIDROID has the ability to analyze, learn, and classify behaviors without the need for any malware data.

The summary of behavioral malware detection techniques described in this section is presented in Table 1.

**Table 1: Comparative behaviors analysis of existing techniques**

| Techniques | Features | Accuracy of detection |
|---|---|---|
| Madam | Behavior monitoring is carried out at the user, application, package, and kernel level. | The detection accuracy is 96.9% on a sample of 2,784 malicious applications, with a low false-positive rate. |
| M0droid | The Framework is composed of an analyzer server and a lightweight client. The analysis and detection are based on the system call invoked by the analyzed application. | 60.16% is the malware detection rate with 0.4% false negatives and 39.43% as a false positive rate. |
| DroidDetector | This hybrid solution is based on a Deep Learning method (Deep Belief Network -DBN-) to extract, investigate, and classify the static and dynamic characteristics of the analyzed application. | With a sample of 1,760 malicious applications, the detection accuracy reached 96.76% |
| Andromaly | For better efficiency, the alert history is combined with the analysis results made on the dynamic characteristics of the monitored applications | Achieve accuracy in between 87% and 99% |
| ServiceMonitor | Static models representing the use of resources, and call systems are made based on the behavior of applications with the different used resources. | Malicious applications are detected with an accuracy rate of 96.7% |
| HIDROID | Learning and classifying behaviors are performed without the use of any malware data. | Detection accuracy reached 0.91 with a false positive rate limited to 0.03. |

Despite the multitude of methods for observing, analyzing, and investigating the applications analyzed, some malicious behaviors can remain undetectable. This can be due to: (1) non-satisfaction of the conditions required for its triggering (Smartphone brand or a specific version of Android for example). (2) the difficulty of identifying similarities between the characteristics previously extracted with those detected during the analysis of a behavior; this is due to the complexity linked to determining clearly and exhaustively the different functionalities of behavior and its associated characteristics. (3) or it is because of the nature of the runtime environment. Indeed, the use of a virtual environment (emulator) can be easily detectable thanks to the methods already implemented in malware. This will give malware the opportunity to mask their malicious behavior and remain undetectable.

## Possible Future Research Prospects

Cybersecurity researchers have the interest of improving the effectiveness of current approaches to detecting malicious applications and especially those interested in behavioral analysis of malware. This type of analysis has shown its effectiveness in several studies; hence the need to concentrate efforts in improving the capacities of this component.

The use of a few features by benign and malicious applications creates some confusion in the classification of these features and increases the false positive rate. To overcome this malfunction and improve detection performance, it is necessary or even essential to add other layers of analysis using other detection techniques (static or heuristic for example) in order to confirm or reverse a decision before communicating it. These layers will have the mission of verifying the suspect functionalities and characteristics, and not redoing a new independent analysis as suggested by some hybrid approaches. This will also help to ensure an adaptation to the different reactions and transformation that malware (normal, polymorphic or other) can undergo depending on the deployment environment (real, virtual device, etc.), the version of Android or even according to the nature of overlay added by the manufacturer.

The approaches based on Deep Learning have given satisfactory results and may also be an area of research that can be explored by experimenting all the algorithms proposed by this method in order to improve the classification of the characteristics and functionalities processed and, thus, improve the performance of the decisions taken.

Having complete coverage of the execution and control paths of an application is considered a major challenge to reveal and block any activity likely to harm the security of devices; and this requires the adoption of a new distributed architecture allowing a better inventory of the different behaviors of the analyzed application. The launching of several simultaneous dynamic analyzes of the latter in different environments will allow the identification and capture of new behaviors (data flow, system calls, etc.). Ouaguid, Abghour, and Ouzzif (2018) had already proposed a similar architecture based on the Blockchain (ANDROSCANREG) but intended only for the static analysis of Android applications. Adopting such an architecture in the dynamic analysis will help improve the detection of malicious behavior which activates if specific conditions are satisfied.

## Conclusion

In this article, we have presented some techniques for detecting malicious applications and we have specifically examined some dynamic approaches based on malware behavioral analysis. These malware are increasingly sophisticated and difficult to detect in the real world despite the integration of Machine Learning and Deep Learning methods in the classification of generic characteristics, rules, and models identifying the behaviors captured.

We believe that the solution to these limits is to have a hybrid method based on a hierarchical combination of different analysis techniques in a distributed environment (based on Blockchain technology for example). Such solution will allow better accuracy in the classification of functionalities extracted during the dynamic analysis of an application or from application samples (benign and malicious) that are analyzed during the training phase of the system. This direction deserves an in-depth study and could open up other opportunities to improve the detection of malicious behavior from more sophisticated malware (polymorphic for example) which changes its representations, its behavior and it is triggered if specific conditions are satisfied.